\documentclass[a4paper]{article}

\usepackage{INTERSPEECH2021}
\usepackage{subfigure}
\usepackage{amsmath}
\usepackage{amsmath}
\usepackage{amsfonts}
\usepackage{amsthm}
\usepackage{subfigure}
\usepackage{appendix}
\usepackage{comment}
\usepackage{multirow}
\usepackage{soul}
\usepackage[table]{xcolor}

\newcommand{\kh}[1]{{#1}}

\newcommand{\method}{Stochastic Future Context}

\newcommand{\cgr}{\cellcolor[gray]{.90}}

\newcommand{\R}{\mathbb{R}}

\title{Multi-mode Transformer Transducer with Stochastic Future Context}

\name{Kwangyoun Kim$^1$, Felix Wu$^1$, Prashant Sridhar$^1$, Kyu J. Han$^1$, Shinji Watanabe$^2$}
\address{
  $^1$ASAPP, USA\\
  $^2$Carnegie Mellon University, USA}
\email{\{kkim,fwu,psridhar,khan\}@asapp.com}

\begin{document}

\maketitle
\begin{abstract}
Automatic speech recognition (ASR) models make fewer errors when more surrounding speech information is presented as context. 
Unfortunately, acquiring a larger future context leads to higher latency. 
There exists an inevitable trade-off between speed and accuracy.
Na\"ively, to fit different latency requirements, people have to store multiple models and pick the best one under the constraints.
Instead, a more desirable approach is to have a single model that can dynamically adjust its latency based on different constraints, which we refer to as \textit{Multi-mode ASR}.
A Multi-mode ASR model can fulfill various latency requirements during inference --- when a larger latency becomes acceptable, the model can process longer future context to achieve higher accuracy and \kh{when a latency budget is not flexible, the model can be less dependent on future context but still achieve reliable accuracy}. In pursuit of Multi-mode ASR, we propose \textit{Stochastic Future Context}, a simple training procedure that samples one streaming configuration in each iteration.
Through extensive experiments on AISHELL-1 and LibriSpeech datasets, we show that a Multi-mode ASR model rivals, if not surpasses, a set of competitive streaming baselines trained with different latency budgets.
\end{abstract}

\section{Introduction} \label{sec:intro}

\kh{End-to-end (E2E) ASR, since its introduction in the early 2000s, has been one of the mainstream research topics in the community. It laid out a path to numerous scientific advances such as Connectionist Temporal Classification (CTC) \cite{graves2006connectionist}, Sequence Transducer \cite{Graves2012SequenceTW}, Neural Transducer \cite{jaitly2015neural}, Listen, Attend and Spell (LAS) \cite{chan2015listen}, etc. Recently, E2E ASR systems like Transformer-Transducer \cite{Zhang2020TransformerTA, yeh2019transformer} keep improving state-of-the-art results on benchmark testsets, especially on LibriSpeech \cite{Panayotov2015LibrispeechAA}. In addition to the development from a research standpoint, this new paradigm is fast transforming the global voice application industry. Now we have more on-device or real-time ASR applications empowered by E2E ASR, thanks to its lower memory footprint and flexibility to meet a number of requirements in production environments for speech recognition systems.}

For streaming \kh{E2E ASR}, various methods such as CTC, Monotonic Attention \cite{raffel2017online}, and Sequence Transducer have been proposed and showed promising results. CTC is a widely known method, and due to \kh{its assumption of conditional independence} between output steps, it is often used with an external Language Model or WFST (Weighted Finite-State Transducer)~\cite{miao2015eesen}. Also it can play an auxiliary role in a joint decoding with other E2E models~\cite{hori-etal-2017-joint, watanabe2017ctcjoint}.
Monotonic attention is a method \kh{inspired by the attention alignment between input speech and output texts being} a monotonic tendency. In particular, MoChA (Monotonic Chunk-wise Attention), which uses both hard-monotonic attention and soft-chunk attention, presents good performance \kh{by enabling attention in streaming scenarios} ~\cite{chiu2018monotonic}.
A sequence transducer normally consists of two encoders to extract latent embedding representations from speech and text sequences. And a joint network is used to combine the two different dimensional embeddings from the encoders. A model is trained through a sequence loss similar to CTC on the combined embedding sequences. Nowadays, sequence transducers have attracted a lot of attention because it has advantages in both accuracy and latency~\cite{chiu2019comparison,weng2020minimum,li2020developing}.

Despite these advancements, streaming E2E ASR still inflicts performance degradation compared to offline models since it doesn't have much freedom of processing future speech frames due to latency constraint. In practice, although varying depending on applications, latency requirements usually sit at around 300ms (median) and less than 1s (95\%-tile). 
In order to solve this challenge, several methods have been studied, especially based on joint training~\cite{narayanan2020cascaded} and knowledge distillation~\cite{Panchapagesan2020EfficientKD}. Recently, a framework called Dual-mode ASR has been introduced, where a single model is trained with two different modes: streaming and full-context~\cite{Yu2020DualmodeAU}. The model trained by this method can operate in both modes, and in particular, the performance of the streaming mode has improved compared to the other studies mostly thanks to inplace knowledge distillation from the offline mode during training.

In this paper, we propose a \textit{Multi-mode ASR} training method that applies a stochastic condition to the dual-mode approach. ASR models are generally trained with a fixed latency budget like 100ms for smartphone applications or 500ms for speech analytics solutions. This would impose accuracy drop on a model trained with one latency specification being used in mismatched latency constraints during inference. To achieve needed accuracy in production environments with various latency requirements, different models specifically optimized to a certain latency condition are required, which would cause a serious scaling issue when serving diverse ASR needs in the industry. Our proposal considering stochastic future contexts can regularize a single model such that it can be exposed to diverse future context settings during training to make it less susceptible to accuracy degradation even in the mismatched inference conditions for latency. Thus, the trained model can serve in any latency budget scenarios without significant WER increase. Experiments are performed based on a Transformer Transducer model~\cite{Zhang2020TransformerTA,yeh2019transformer}, which uses a Transformer layer in encoders. We present the advantages of the proposed method through comparisons and analysis with the existing methods.

\section{Related works} \label{sec:related_works}

\begin{figure*}[!t]
    \centering
    \subfigure[]{
    \includegraphics[width=0.3\linewidth]{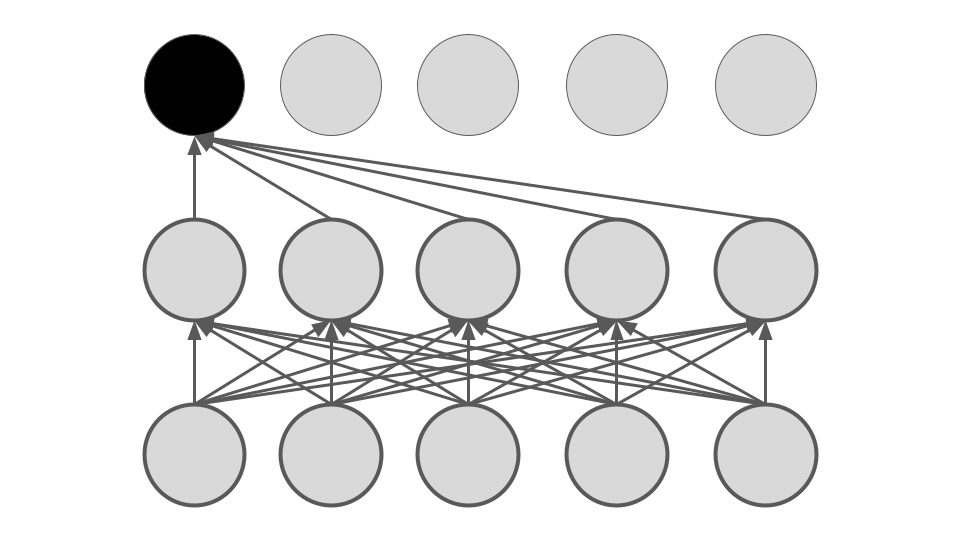}
    }
    \subfigure[]{
    \includegraphics[width=0.3\linewidth]{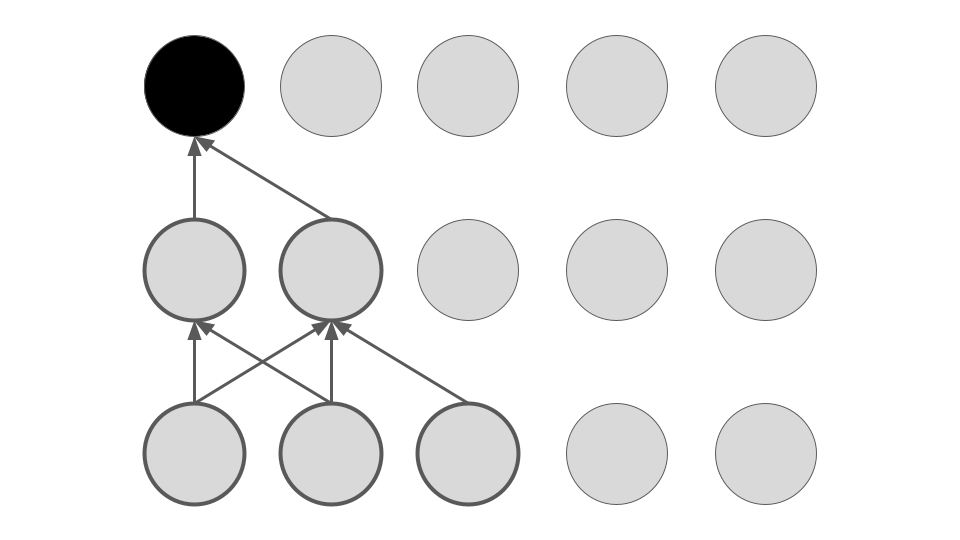}
    }
    \subfigure[]{
    \includegraphics[width=0.3\linewidth]{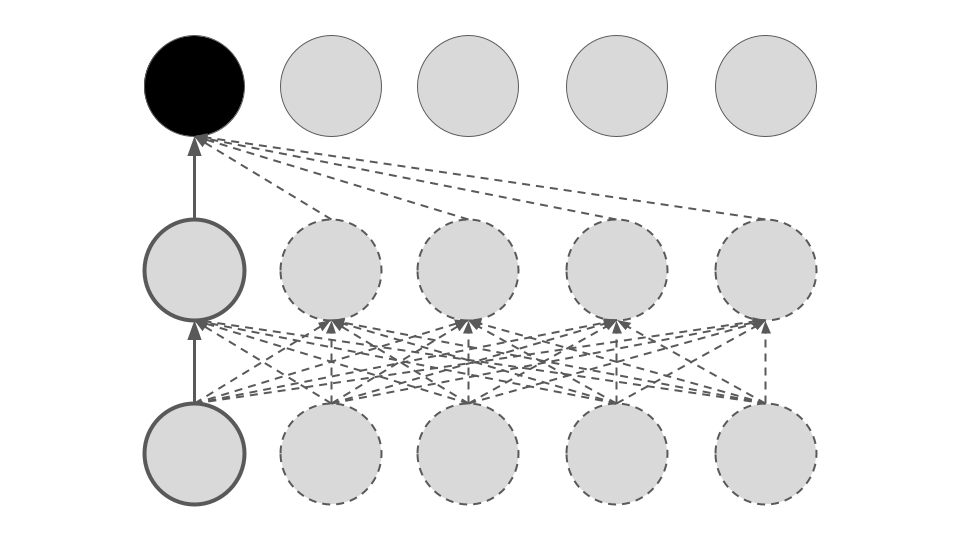}
    }
    \vspace{-0.3cm}
    \caption{A figure of different modes. The black circle indicates the current output step, and bold-lined circles are contexts used for the current output. (a) is a full-context mode, and (b) presents a streaming mode with the future context size of 1. (c) describes our method which randomly selects the future context (shown with dotted circles and arrows).}
    \label{fig:modes}
    \vspace{-0.2cm}
\end{figure*}

\subsection{Sequence Transducers} 

Sequence Transducers  \cite{Graves2012SequenceTW,Zhang2020TransformerTA,yeh2019transformer,Gulati2020ConformerCT,Tripathi2020TransformerTO} is a popular framework for both offline and streaming ASR, which achieves state-of-the-art results recently on LibriSpeech~\cite{Zhang2020PushingTL}. 
A sequence transducer ASR model is composed of three modules: an audio encoder, an autoregressive label encoder, and a joint network.
The audio encoder takes either raw waveform or a sequence of mel-spectral features $\{\mathbf{x}_1, \mathbf{x}_2, \dots, \mathbf{x}_N\}$ as inputs and produces a sequence of embeddings $\{\mathbf{h}^a_1, \mathbf{h}^a_2, \dots, \mathbf{h}^a_T\}$. 
The audio encoder usually downsamples the input sequence which makes $T < N$. 
The label encoder takes all previous tokens $\{y_1, y_2, \dots, y_{u-1}\}$ and generates $\mathbf{h}^l_{u}$. 
The joint network uses the embeddings from the both encoders $\mathbf{h}^a_{t}$ and $\mathbf{h}^l_{u}$ and generates $\mathbf{h}^j_{t, u}$ as 
\begin{equation}
\mathbf{h}^j_{t,u}=\bm{W}\tanh({\bm{W}}^a\mathbf{h}^a_{t}+{\bm{W}}^l\mathbf{h}^l_{u})
\end{equation}
The output probability is computed through the softmax as, ${Pr(k \in \bar{Y}} | t, u) = \mathbf{P}_{t,u} = \frac{\exp{\mathbf{h}^j_{t,u}}}{\sum_{u'}\exp{\mathbf{h}^j_{t,u'}}}$, where $\bar{Y}$ is the set of token whose size is $V$ including the blank token $\phi$.

Traditionally, both of the audio and label encoders are comprised of RNNs and the joint network is simply a two-layer feed-forward network. Recently, Transformers~\cite{Zhang2020TransformerTA,yeh2019transformer,Tripathi2020TransformerTO} and Conformers~\cite{Gulati2020ConformerCT} have been shown to be more competitive alternatives to RNNs as an encoder architecture. 

\subsection{Dual-mode ASR}
Dual-mode ASR~\cite{Yu2020DualmodeAU} is a framework to unify streaming and full-context (offline) models through weight-sharing.
While using the same trained weights, the model can decode in either full-context or streaming mode based on whether the whole context is available. As shown in Fig ~\ref{fig:modes}, Dual-mode ASR uses a pair of the full-context mode as in $(a)$ and the streaming mode with a fixed future context size as in $(b)$.
In the streaming mode, the model masks out the future context in its self-attention and convolution layers.
Na\"ively sharing the parameters of the two modes can lead to performance degradation. To address this issue, Yu et al.~\cite{Yu2020DualmodeAU} proposed to employ joint training and inplace knowledge distillation. With these two techniques, the model is optimized with the following loss in each training iteration:
\begin{align}
    L_{\text{dual}} = & L_{\text{trans}}(\mathbf{P}^0, \mathbf{y}) + L_{\text{trans}}(\mathbf{P}^{\infty}, \mathbf{y}) +
    L_{\text{distil}}(\mathbf{P}^{0}, \mathbf{P}^{\infty}),
    \label{eq:total_loss}
\end{align}
where $\mathbf{y}$ is the ground-truth sequence of tokens, whose length is $U$, $\mathbf{P}^{0}$ and $\mathbf{P}^{\infty} \in \R^{U \times V}$ are the streaming mode and full-context mode outputs, respectively,
$L_{\text{trans}}(\cdot, \cdot)$ is the transducer loss and $ L_{\text{distil}}(\cdot, \cdot)$ is the inplace knowledge distillation loss.
Instead of taking the direct KL-divergence between $\mathbf{P}^{0}$ and $\mathbf{P}^{\infty}$, Dual-mode ASR follows \cite{Panchapagesan2020EfficientKD} to merge the probabilities of unimportant tokens for the efficient knowledge distillation
\begin{equation}
    \hat{\mathbf{P}}^{\text{0}}_{u, :} = \left(
    \mathbf{P}^{\text{0}}_{\phi};  \mathbf{P}^{\text{0}}_{y_{u}}; \mathbf{1} - \mathbf{P}^{\text{0}}_{\phi} - \mathbf{P}^{\text{0}}_{y_{u}}  \right) 
\end{equation}
and computes the KL-divergence between  $\mathbf{\hat{P}}^{0}$ and $\mathbf{\hat{P}}^{\infty}$ with a time shift of $s$ steps. This time shifting compensates the difference of the frame alignment between the both modes, since the full-context mode usually has less emission latency compared to the streaming mode. Precisely, the distillation loss is defined as
\begin{align}
    L_{\text{distil}}(\mathbf{P}^{0}, \mathbf{P}^{\infty}) = D_{\text{KL}}(\mathbf{\hat{P}}^{0}_{s:T} || \mathbf{\hat{P}}^{\infty}_{1:T-s} ).
\end{align}
A simple extension of the original Dual-mode ASR is to use a small future context $C$ in the audio encoder. This increases the latency of the streaming mode, but significantly reduces the errors. $\mathbf{P}^0$ would be replaced by $\mathbf{P}^C$ in Eq. \ref{eq:total_loss}.
\section{Multi-Mode ASR} \label{sec:method}

In Dual-Mode ASR~\cite{Yu2020DualmodeAU}, the streaming mode is not allowed to use any future context. However, in the real-world scenario where a small amount of latency is still acceptable within a given latency budget depending on applications, using a few frames in the future context is significantly beneficial to improve  accuracy~\cite{Zhang2020TransformerTA,yeh2019transformer,Tripathi2020TransformerTO}.  

Merely having one model with a specified future context, however, would not be the best solution. 
Different speech applications would impose different latency requirements for ASR models. The acceptable latency may vary depending on user perception, device or server status, etc. Every latency constraint would enforce another expensive model training. Hence, a model that supports multiple latency modes without accuracy degradation is critical from a scalable ASR perspective. For similar purpose, a unified ASR system combining online and offline models has been introduced that trains a system with a method called Dynamic Latency Training (DLT)~\cite{gao2020universal}.

\subsection{\method}
To enable Multi-mode ASR, we propose \method{} --- instead of using a fixed value, the future context size $C$ of the streaming mode is drawn from a random distribution $\tilde{P}$ during training.
This encourages the audio encoder to learn to utilize various amount of future information.
Admittedly, an adequate distribution $\tilde{P}$ should be chosen according to use cases.

For simplicity, we keep the full-context mode as is so it can still serve as a teacher for inplace knowledge distillation.
Similar to Eq. \ref{eq:total_loss}, Multi-mode ASR models are trained with the following objective:
\begin{multline}
    L_{\text{multi}} =
    E_{C \sim \tilde{P}} \left[ L_{\text{trans}}(\mathbf{P}^{C}, \mathbf{y}) + L_{\text{trans}}(\mathbf{P}^{\infty}, \mathbf{y}) \right.\\
    \left.+ L_{\text{distil}}(\mathbf{P}^{C}, \mathbf{P}^{\infty}) \right],
    \label{eq:total_loss_multi}
\end{multline}
where $\mathbf{P}^{C}$ is the output probability of the streaming mode with the stochastic future context size $C \sim \tilde{P}$. 

For applying the stochastic size to the Transformer of the model, we mainly use three different methods.  

\subsubsection{Tied Mask across Layers}
We sample a single value $c$ from a chosen distribution per mini-batch and set the future context size of all layers as $c$.
This makes the total future context size $C = c \cdot L$, where $L$ is the number of self-attention layers in a Transformer encoder.
To illustrate, $C$ can be $0, L, 2L, 3L$, etc.

\subsubsection{Untied Mask across Layers}
For each layer $l$, we sample the future context size $c_l$ independently. The total right context size $C = \sum_{l} c_l$.
This makes $C$ more diverse since it is not always a multiple of $L$. To illustrate, $C$ can be $0, 1, 2, 3$, etc.

\subsubsection{Untied Mask under a Constraint}
We add a constraint of the total future context size on the untied mask method described in Section 3.1.2. A model learns multiple modes for each layers but the total future context size is always below  $C$, which is predefined. 
Under the constraint, a future context size for each layers is sampled iteratively. In order to avoid too much concentration at the bottom layers when assigning future contexts, an adjusted uniform distribution is used for sampling as follows,
\begin{equation}
c \sim \mathcal{U}(0, R/d)
\end{equation}
 where $R$ is a remaining future context size under the constraint, and $d$ is a tuned parameter.
\section{Experiments}
We use ESPnet~\cite{watanabe2018espnet} to train and test models in our experiments. Several related parts have been implemented internally or integrated with Fairseq~\cite{ott2019fairseq}.
Also we use NVIDIA Apex AMP (Automatic Mixed Precision) with the optimization level of O2.
Our Transformer Transducer model has 12 audio encoder and 2 label encoder layers.
Each Transformer layer has a 512-dim embedding, a feed-forward with the hidden layer size of 2048, 8 heads for self-attention, and the attention dropout rate of 0.1 similar to the Transformer-base~\cite{Vaswani2017AttentionIA}.
Following \cite{Radford2018ImprovingLU,Devlin2019BERTPO}, we use GELU activations~\cite{hendrycks2020gaussian}  in the feed-forward modules.
Two 2D strided convolution layers are applied at the bottom of the audio encoder, and each convolution layer has 32 channels, (3$\times$3) kernel and (2$\times$2) stride.
The hidden layer size of the joint network is $256$.
We use 80-dim filterbank features computed over a 25ms window and a 10ms shift as inputs.
We apply adaptive SpecAugment~\cite{Park_2019,Park2020SpecaugmentOL} in all experiments. 3-best checkpoints during training epochs are selected in terms of the lowest validation loss, and their average is used as the final model.
In our experiments, two stochastic distributions are used for the proposed Multi-mode ASR training, i.e., discrete uniform distribution $\mathcal{U}$ and normal distribution $\mathcal{N}$. Because the normal distribution is continuous and spans over negative and postive values, we use a \texttt{floor} and \texttt{abs} function to discretize absolute values. The reason for using the normal distribution instead of a multinomial distribution is that it is easy to configure by adjusting only two parameters (mean and variance).
We evaluate our methods on two public datasets: AISHELL-1 and LibriSpeech.

\subsection{AISHELL-1}
AISHELL-1~\cite{Bu2017AISHELL1AO} is an open-sourced Mandarin speech recognition corpus. It contains 150h, 10h, and 5h of audio for training, development, and testing, respectively. We use characters as output tokens, and the total number of tokens is about 4K including special tags such as \textless blank\textgreater and \textless unknown\textgreater. The speed perturbation is applied with factors of 0.9, 1.0 and 1.1 to augment data. The size of mini-bathes is set to 96000, a learning rate is set to 0.0005, and a learning rate warm-up is enabled with 10K warm-up steps.

\begin{table*}[th]
  \caption{CER on AISHELL-1 with different future context sizes applied on a single model during inference. The gray field indicates a mismatched condition when the future context sizes used for training and recognition are different.}
  \vspace{-0.1cm}
  \label{tab:aishell_all}
  \centering
  \begin{tabular}{cccccccccc}
  \toprule
  & Future & \multicolumn{8}{c}{Future Context Size (at Inference)} \\
  & Context Size& \multicolumn{2}{c}{c = 0} & \multicolumn{2}{c}{c = 1} & \multicolumn{2}{c}{c = 2} & \multicolumn{2}{c}{c = $\infty$ (\text{Full Context})}\\
  & (at Training) & dev & test & dev & test & dev & test & dev & test \\
  \hline\hline
  \multirow{3}{*}{Baseline} 
  & 0 & 8.23\% & 9.14\% & \cgr12.55\% & \cgr13.37\% & \cgr58.07\% & \cgr59.55\% & \cgr141.19\% & \cgr139.95\% \\
  & 1 & \cgr53.76\% & \cgr54.46\% & 8.23\% & 9.11\% & \cgr9.37\% & \cgr10.25\% & \cgr109.93\% & \cgr115.46\% \\
  \hline
  \multirow{3}{*}{Dual-mode} 
  & c = 0 & 7.74\% & 8.51\% & \cgr7.66\% & \cgr8.44\% & \cgr7.62\% & \cgr8.41\% & 6.63\% & 7.23\% \\
  & c = 1 & \cgr23.46\% & \cgr25.07\% & 7.64\% & 8.30\% & \cgr7.51\% & \cgr8.16\% & 6.79\% & 7.23\% \\
  & c = 2 & \cgr31.83\% & \cgr34.18\% & \cgr8.38\% & \cgr9.07\% & 6.95\% & 7.59\% & 6.41\% & 6.86\% \\
  \hline
  \multirow{8}{*}{Tied mask} 
  & $c \sim \mathcal{U}(0, 1)$ & 8.01\% & 8.62\% & 7.58\% & 8.20\% & 7.49\% & 7.97\% & 6.68\% & 7.20\%\\
  & $c \sim \mathcal{U}(0, 2)$ & 7.73\% & 8.52\% & 7.20\% & 8.05\% & 7.01\% & 7.74\% & 6.41\% & 6.99\%\\
  & $c \sim \mathcal{U}(0, 10)$ & 8.91\% & 9.73\% & 7.32\% & 8.12\% & 6.80\% & 7.60\% & 6.13\% & 6.73\% \\
  \cline{2-10}
  & $c \sim \mathcal{N}(0, 0.5)$ & 7.57\% & 8.39\% & 7.49\% & 8.27\% & 7.42\% & 8.16\% & 6.60\% & 7.30\% \\
  & $c \sim \mathcal{N}(0, 1)$ & 7.84\% & 8.49\% & 7.66\% & 8.29\% & 7.53\% & 8.10\% & 6.80\% & 7.24\% \\
  & $c \sim \mathcal{N}(0, 2)$ & 7.79\% & 8.60\% & 7.31\% & 8.08\% & 7.08\% & 7.91\% & 6.40\% & 7.17\% \\
  & $c \sim \mathcal{N}(1, 1)$ & 7.95\% & 8.84\% & 7.27\% & 8.10\% & 7.04\% & 7.85\% & 6.48\% & 7.17\% \\
  \hline
  \multirow{3}{*}{Untied mask} 
  & $c \sim \mathcal{U}(0, 2)$ & 8.91\% & 9.81\% & 7.37\% & 8.06\% & 7.20\% & 7.87\% & 6.58\% & 7.07\% \\
  & $c \sim \mathcal{U}(0, 10)$ & 15.56\% & 17.36\% & 8.59\% & 9.71\% & 7.20\% & 8.00\% & 6.01\% & 6.53\% \\
  & $c \sim \mathcal{N}(0, 0.5)$ & 7.78\% & 8.41\% & 7.65\% & 8.24\% & 7.63\% & 8.21\% & 6.74\% & 7.23\% \\
  \hline
  Constraint& $C=12$ & 8.66\% & 9.62\% & 7.46\% & 8.18\% & 7.32\% & 8.04\% & 6.58\% & 7.14\% \\
  \bottomrule
  \vspace{-0.5cm}
  \end{tabular}
\end{table*}

\subsection{LibriSpeech}
LibriSpeech~\cite{Panayotov2015LibrispeechAA} is an open-sourced English speech recognition corpus extracted from audiobooks. It contains 960h of training audios and a disjoint text-only corpus of 800M word tokens for language model training. It has two development sets and two test sets: dev-clean, dev-other, test-clean, and test-other. Each of them has about 5h of audio. We train models with 1K BPE sub-word units as output tokens, use 328000 as the size of mini-batches and set a learning rate to 0.001. We apply a learning rate scheduler as ~\cite{Park_2019} to divide training steps into 3 stages, i.e., 13K warm-up, 40K holding and 77K exponential decay steps. 

\begin{table}[th]
    \caption{Streaming WER on LibriSpeech testset with different future context sizes applied on a single model during inference. $c$ is a future context size at training. Note that the results from the literature confirm our baselines are within a reasonable range, but are not used for direct comparison.}
    \vspace{-0.1cm}
  \label{tab:librispeech_diff}
  \centering
  \begin{tabular}{lcccc}
  \toprule
  & \multicolumn{4}{c}{Future Context Size (at Inference)} \\
  & \multicolumn{2}{c}{c = 0} & \multicolumn{2}{c}{c = 1} \\
  & (test)clean & other & clean & other\\
  \hline\hline
  Other works & & & & \\
  \multicolumn{1}{r}{Yu et al.~\cite{Yu2020DualmodeAU}} & 5.0\% & 11.6\% & - & - \\
  \multicolumn{1}{r}{Yeh et al.~\cite{yeh2019transformer}} & 12.32\% & 23.08\% & 6.99\% & 16.88\% \\
  \hline
  Baseline & & & & \\
  \multicolumn{1}{r}{$c=0$} & 7.09\% & 15.56\% & \cgr7.70\% & \cgr16.39\% \\
  \multicolumn{1}{r}{$c=1$} & \cgr55.51\% & \cgr70.72\% & 6.37\% & 13.77\%\\
  \hline
  Dual-mode & & & & \\
  \multicolumn{1}{r}{$c=0$} & 7.24\% & 15.07\% & \cgr7.13\% & \cgr15.16\% \\
  \multicolumn{1}{r}{$c=1$} & \cgr42.81\% & \cgr59.90\% & 6.12\% & 13.46\% \\
  \hline
  Tied mask & & \\
  \multicolumn{1}{r}{$c \sim \mathcal{U}(0, 3)$} & 7.16\% & 15.61\% & 6.15\% & 13.23\% \\
  \multicolumn{1}{r}{$c \sim \mathcal{N}(0, 2)$} & 7.16\% & 15.11\% & 6.19\% & 13.35\% \\
  \hline
  Untied mask & & \\
  \multicolumn{1}{r}{$c \sim \mathcal{U}(0, 3)$} & 9.78\% & 19.73\% & 6.33\% & 13.90\% \\
  \hline
  Constraint & & \\
  \multicolumn{1}{r}{$C=12$} & 7.94\% & 17.31\% & 6.09\% & 13.55\% \\
  \bottomrule
  \vspace{-0.8cm}
  \end{tabular}
\end{table}
\section{Discussion}
We employ a 12 layer Transformer in the audio encoder with a time-shift of 40ms due to convolutions, so inference with the future context size of 1 for each layer would enforce 480ms latency. Considering less than 1s is practically acceptable latency for many applications, we consider the streaming mode with two or less future context size per layer ($c \le 2$ or $C \le 12$) in the discussion. 

Table ~\ref{tab:aishell_all} shows the CERs (Character Error Rates) of various models with different methods and future context sizes on AISHELL-1. While baseline streaming models are trained as stand-alone, dual-mode and multi-mode models learn from a pair of a full-context mode and a streaming mode with a specified or randomly chosen context size. First, the overall result shows that the baselines usually operate well only at the same context size used in training (i.e., matched condition). In other cases, the performance drops a lot (i.e., mismatched condition, highlighted as gray). The dual-mode models seem to work reasonably at future context sizes larger than that used for training. We argue that the reason is the models trained by the dual-mode approach learn those modes with larger future contexts indirectly from the full-context mode. However, in the cases of smaller context sizes than those specified for training, we can see noticeable performance degradation. For example, when we see the results of the dual-mode model trained with $c = 1$, there is a difference of more than 15\% (absolute) when this model performs at $c = 0$ and 1 (i.e., 23.46\% vs. 7.64\% for dev and 25.07\% vs. 8.30\% for test). 

In contrast, the multi-mode ASR models perform well in various streaming modes in general, showing resilience to even mismatched conditions in future context setting. Probability distributions used in multi-mode experiments are set to select what is more often in the crucial range of context. As shown in the tied mask results, frequently selected future context mode works better than other modes. For example, in the case of using discrete uniform distributions to decide a stochastic context size, $c \sim \mathcal{U}(0,1)$ outperforms $c \sim \mathcal{U}(0,10)$ at $c = 0$ at inference, but when the size becomes larger like 2, it does not. On the other hand, the results in the case of normal distributions show other aspects. For example, $c \sim \mathcal{N}(0,1)$ and $c \sim \mathcal{N}(1,1)$ differ only in mean, and they outperform each other in the mode matched with their mean. When analyzing $c \sim \mathcal{N}(0, \cdot) $ series, we can find that the model trained with a larger standard deviation works better at larger future context sizes. These results prove that the multi-mode performs well in the overall mode, and show that the distribution can be set according to practical purposes.

With untied masks, the model behaves differently even with the same distribution.
First, as described in Section 3, the untied mask method trains a model with the future context size $C \in \{0, 1, 2, ...\}$, while the tied mask method uses $C \in \{0, L, 2L\}$, etc. So, with the united mask method, the frequency for each context size to be selected and shown during training relatively decreases. For example, when $c \sim \mathcal{U}(0,10)$ is used with the untied mask method, this issue is seen noticeable to make the performance degradation at the context size 0. Conversely, in the case of the full-context, the performance is better than other methods. This can be analyzed that this is regularization effect from much more diversely applied future context masks for each layer. When we add a constraint of the desired future context size as 12 to the untied mask method to address the aforementioned issue, the performance degradation at the future context 0 mode is partially solved. However, the overall accuracy is lower than the tied mask method.  

Table~\ref{tab:librispeech_diff} shows the LibriSpeech results.
We observe a similar trend as AISHELL-1.
With the tied mask $c \sim \mathcal{N}(0, 2)$,
a multi-mode ASR model shows better or similar WER to two dual-model models each trained in matched conditions. To make a clear comparison, we compare the average WER of each model at the future context sizes of 0 and 1, which are practical cases for the streaming ASR. The averaged WER of the dual-mode models trained with $c = 0$ is 7.19\%, 15.11\% on test-clean and test-other, respectively. But, the multi-mode model trained with $c \sim \mathcal{N}(0,2)$ surpasses it with 6.68\%, 14.23\% on the same test sets. Based on this, we find that the multi-mode ASR model outperforms others overall in multiple latency budget situations, especially in critical cases for practical streaming ASR.
\section{Conclusions}

Since a single model is trained as multi-mode, we expect the trained model performs well in multiple future context modes. In other words, it is possible to dynamically adjust the future context size without any additional training. Through the experimental results, we verified that our proposed method can has this flexibility. It is also expected that this proposed method can be used for practical purposes such as a dynamic delay management and load balancing in an actual speech recognition service. Presumably, further fine-tuning a Multi-mode ASR model on the test configurations can achieve higher accuracy; however, we leave this as future work.

\bibliographystyle{IEEEtran}

\bibliography{mybib}

\begin{thebibliography}{10}
\providecommand{\url}[1]{#1}
\csname url@samestyle\endcsname
\providecommand{\newblock}{\relax}
\providecommand{\bibinfo}[2]{#2}
\providecommand{\BIBentrySTDinterwordspacing}{\spaceskip=0pt\relax}
\providecommand{\BIBentryALTinterwordstretchfactor}{4}
\providecommand{\BIBentryALTinterwordspacing}{\spaceskip=\fontdimen2\font plus
\BIBentryALTinterwordstretchfactor\fontdimen3\font minus
  \fontdimen4\font\relax}
\providecommand{\BIBforeignlanguage}[2]{{%
\expandafter\ifx\csname l@#1\endcsname\relax
\typeout{** WARNING: IEEEtran.bst: No hyphenation pattern has been}%
\typeout{** loaded for the language `#1'. Using the pattern for}%
\typeout{** the default language instead.}%
\else
\language=\csname l@#1\endcsname
\fi
#2}}
\providecommand{\BIBdecl}{\relax}
\BIBdecl

\bibitem{graves2006connectionist}
A.~Graves, S.~Fern{\'a}ndez, F.~Gomez, and J.~Schmidhuber, ``Connectionist
  temporal classification: labelling unsegmented sequence data with recurrent
  neural networks,'' in \emph{Proceedings of the 23rd international conference
  on Machine learning}, 2006, pp. 369--376.

\bibitem{Graves2012SequenceTW}
A.~Graves, ``Sequence transduction with recurrent neural networks,''
  \emph{ArXiv}, vol. abs/1211.3711, 2012.

\bibitem{jaitly2015neural}
N.~Jaitly, D.~Sussillo, Q.~V. Le, O.~Vinyals, I.~Sutskever, and S.~Bengio, ``A
  neural transducer,'' \emph{arXiv preprint arXiv:1511.04868}, 2015.

\bibitem{chan2015listen}
W.~Chan, N.~Jaitly, Q.~V. Le, and O.~Vinyals, ``Listen, attend and spell,''
  \emph{arXiv preprint arXiv:1508.01211}, 2015.

\bibitem{Zhang2020TransformerTA}
Q.~Zhang, H.~Lu, H.~Sak, A.~Tripathi, E.~McDermott, S.~Koo, and S.~Kumar,
  ``Transformer transducer: A streamable speech recognition model with
  transformer encoders and rnn-t loss,'' \emph{ICASSP 2020 - 2020 IEEE
  International Conference on Acoustics, Speech and Signal Processing
  (ICASSP)}, pp. 7829--7833, 2020.

\bibitem{yeh2019transformer}
C.-F. Yeh, J.~Mahadeokar, K.~Kalgaonkar, Y.~Wang, D.~Le, M.~Jain, K.~Schubert,
  C.~Fuegen, and M.~L. Seltzer, ``Transformer-transducer: End-to-end speech
  recognition with self-attention,'' \emph{arXiv preprint arXiv:1910.12977},
  2019.

\bibitem{Panayotov2015LibrispeechAA}
V.~Panayotov, G.~Chen, D.~Povey, and S.~Khudanpur, ``Librispeech: An asr corpus
  based on public domain audio books,'' \emph{2015 IEEE International
  Conference on Acoustics, Speech and Signal Processing (ICASSP)}, pp.
  5206--5210, 2015.

\bibitem{raffel2017online}
C.~Raffel, M.-T. Luong, P.~J. Liu, R.~J. Weiss, and D.~Eck, ``Online and
  linear-time attention by enforcing monotonic alignments,'' in
  \emph{International Conference on Machine Learning}.\hskip 1em plus 0.5em
  minus 0.4em\relax PMLR, 2017, pp. 2837--2846.

\bibitem{miao2015eesen}
Y.~Miao, M.~Gowayyed, and F.~Metze, ``Eesen: End-to-end speech recognition
  using deep rnn models and wfst-based decoding,'' in \emph{2015 IEEE Workshop
  on Automatic Speech Recognition and Understanding (ASRU)}.\hskip 1em plus
  0.5em minus 0.4em\relax IEEE, 2015, pp. 167--174.

\bibitem{hori-etal-2017-joint}
T.~Hori, S.~Watanabe, and J.~Hershey, ``Joint {CTC}/attention decoding for
  end-to-end speech recognition,'' in \emph{Proceedings of the 55th Annual
  Meeting of the Association for Computational Linguistics (Volume 1: Long
  Papers)}, 2017.

\bibitem{watanabe2017ctcjoint}
S.~Watanabe, T.~Hori, S.~Kim, J.~R. Hershey, and T.~Hayashi, ``Hybrid
  ctc/attention architecture for end-to-end speech recognition,'' in \emph{IEEE
  Journal of Selected Topics in Signal Processing, (Volume:11, Issue:8)}, 2017.

\bibitem{chiu2018monotonic}
C.-C. Chiu and C.~Raffel, ``Monotonic chunkwise attention,'' \emph{arXiv
  preprint arXiv:1712.05382}, 2017.

\bibitem{chiu2019comparison}
C.-C. Chiu, W.~Han, Y.~Zhang, R.~Pang, S.~Kishchenko, P.~Nguyen, A.~Narayanan,
  H.~Liao, S.~Zhang, A.~Kannan \emph{et~al.}, ``A comparison of end-to-end
  models for long-form speech recognition,'' in \emph{2019 IEEE Automatic
  Speech Recognition and Understanding Workshop (ASRU)}.\hskip 1em plus 0.5em
  minus 0.4em\relax IEEE, 2019, pp. 889--896.

\bibitem{weng2020minimum}
C.~Weng, C.~Yu, J.~Cui, C.~Zhang, and D.~Yu, ``Minimum bayes risk training of
  rnn-transducer for end-to-end speech recognition,'' \emph{Proc. Interspeech
  2020}, pp. 966--970, 2020.

\bibitem{li2020developing}
J.~Li, R.~Zhao, Z.~Meng, Y.~Liu, W.~Wei, S.~Parthasarathy, V.~Mazalov, Z.~Wang,
  L.~He, S.~Zhao \emph{et~al.}, ``Developing rnn-t models surpassing
  high-performance hybrid models with customization capability,'' \emph{Proc.
  Interspeech 2020}, pp. 3590--3594, 2020.

\bibitem{narayanan2020cascaded}
A.~Narayanan, T.~N. Sainath, R.~Pang, J.~Yu, C.-C. Chiu, R.~Prabhavalkar,
  E.~Variani, and T.~Strohman, ``Cascaded encoders for unifying streaming and
  non-streaming asr,'' \emph{arXiv preprint arXiv:2010.14606}, 2020.

\bibitem{Panchapagesan2020EfficientKD}
S.~Panchapagesan, D.~Park, C.-C. Chiu, Y.~Shangguan, Q.~Liang, and
  A.~Gruenstein, ``Efficient knowledge distillation for rnn-transducer
  models,'' \emph{ArXiv}, vol. abs/2011.06110, 2020.

\bibitem{Yu2020DualmodeAU}
J.~Yu, W.~Han, A.~Gulati, C.-C. Chiu, B.~Li, T.~N. Sainath, Y.~Wu, and R.~Pang,
  ``Dual-mode asr: Unify and improve streaming asr with full-context
  modeling,'' \emph{Proceedings of ICLR}, 2021.

\bibitem{Gulati2020ConformerCT}
A.~Gulati, J.~Qin, C.-C. Chiu, N.~Parmar, Y.~Zhang, J.~Yu, W.~Han, S.~Wang,
  Z.~Zhang, Y.~Wu, and R.~Pang, ``Conformer: Convolution-augmented transformer
  for speech recognition,'' \emph{ArXiv}, vol. abs/2005.08100, 2020.

\bibitem{Tripathi2020TransformerTO}
A.~Tripathi, J.~Kim, Q.~Zhang, H.~Lu, and H.~Sak, ``Transformer transducer: One
  model unifying streaming and non-streaming speech recognition,''
  \emph{ArXiv}, vol. abs/2010.03192, 2020.

\bibitem{Zhang2020PushingTL}
Y.~Zhang, J.~Qin, D.~Park, W.~Han, C.-C. Chiu, R.~Pang, Q.~V. Le, and Y.~Wu,
  ``Pushing the limits of semi-supervised learning for automatic speech
  recognition,'' \emph{ArXiv}, vol. abs/2010.10504, 2020.

\bibitem{gao2020universal}
Z.~Gao, S.~Zhang, M.~Lei, and I.~McLoughlin, ``Universal {ASR:} unifying
  streaming and non-streaming {ASR} using a single encoder-decoder model,''
  \emph{arXiv preprint arXiv:2010.14099}, 2020.

\bibitem{watanabe2018espnet}
S.~Watanabe, T.~Hori, S.~Karita, T.~Hayashi, J.~Nishitoba, Y.~Unno, N.~{Enrique
  Yalta Soplin}, J.~Heymann, M.~Wiesner, N.~Chen, A.~Renduchintala, and
  T.~Ochiai, ``{ESPnet}: End-to-end speech processing toolkit,'' in
  \emph{Proceedings of Interspeech}, 2018.

\bibitem{ott2019fairseq}
M.~Ott, S.~Edunov, A.~Baevski, A.~Fan, S.~Gross, N.~Ng, D.~Grangier, and
  M.~Auli, ``fairseq: A fast, extensible toolkit for sequence modeling,'' in
  \emph{Proceedings of NAACL-HLT 2019: Demonstrations}, 2019.

\bibitem{Vaswani2017AttentionIA}
A.~Vaswani, N.~Shazeer, N.~Parmar, J.~Uszkoreit, L.~Jones, A.~N. Gomez,
  L.~Kaiser, and I.~Polosukhin, ``Attention is all you need,'' \emph{ArXiv},
  vol. abs/1706.03762, 2017.

\bibitem{Radford2018ImprovingLU}
A.~Radford, K.~Narasimhan, T.~Salimans, and I.~Sutskever, ``Improving language
  understanding by generative pre-training,'' 2018.

\bibitem{Devlin2019BERTPO}
J.~Devlin, M.-W. Chang, K.~Lee, and K.~Toutanova, ``Bert: Pre-training of deep
  bidirectional transformers for language understanding,'' in \emph{NAACL-HLT},
  2019.

\bibitem{hendrycks2020gaussian}
D.~Hendrycks and K.~Gimpel, ``Gaussian error linear units (gelus),''
  \emph{arXiv preprint arXiv:1606.08415}, 2016.

\bibitem{Park_2019}
D.~S. Park, W.~Chan, Y.~Zhang, C.-C. Chiu, B.~Zoph, E.~D. Cubuk, and Q.~V. Le,
  ``Specaugment: A simple data augmentation method for automatic speech
  recognition,'' \emph{Interspeech 2019}, 2019.

\bibitem{Park2020SpecaugmentOL}
D.~Park, Y.~Zhang, C.-C. Chiu, Y.~Chen, B.~Li, W.~Chan, Q.~V. Le, and Y.~Wu,
  ``Specaugment on large scale datasets,'' \emph{ICASSP 2020 - 2020 IEEE
  International Conference on Acoustics, Speech and Signal Processing
  (ICASSP)}, pp. 6879--6883, 2020.

\bibitem{Bu2017AISHELL1AO}
H.~Bu, J.~Du, X.~Na, B.~Wu, and H.~Zheng, ``Aishell-1: An open-source mandarin
  speech corpus and a speech recognition baseline,'' \emph{2017 20th Conference
  of the Oriental Chapter of the International Coordinating Committee on Speech
  Databases and Speech I/O Systems and Assessment (O-COCOSDA)}, pp. 1--5, 2017.

\end{thebibliography}

\end{document}